\documentclass[12pt]{iopart}
\usepackage{amssymb}
\usepackage[amssymb]{SIunits}
\usepackage{graphicx}
\usepackage{xspace}
\usepackage{bm}
\usepackage[colorlinks,urlcolor=blue]{hyperref}
\usepackage[all]{hypcap}
\newcommand*{\doi}[1]{\href{http://dx.doi.org/\detokenize{#1}}{\detokenize{#1}}}

\begin{document}

\title{Particle swarm optimization of divertor targets for heat load control}

\author{H. Frerichs${}^1$}

\address{${}^1$ Department of Nuclear Engineering \& Engineering Physics, University of Wisconsin - Madison, WI, USA}

\ead{hfrerichs@wisc.edu}

\begin{abstract}
Divertor targets in magnetic confinement fusion devices must be designed to handle extreme heat loads.
Fast approximation of heat loads with FLARE based on field line reconstruction from an unstructured flux tube mesh is utilized in particle swarm optimization (PSO) of the divertor target geometry.
Optimization of the outer divertor target in ITER is evaluated with a constraint for the head loads onto baffles.
The optimal configuration is found to depend on assumptions for the background plasma in the heat load proxy simulation.
\end{abstract}

\vspace{2pc}

\def\vec#1{\ensuremath{{\bf #1}}\xspace}
\def\Poincare{Poincar\'e\xspace}
\def\phimap{\ensuremath{\varphi_{\textnormal{map}}}\xspace}
\def\Dmax{\ensuremath{\Delta_{\Phi \, \textnormal{max}}}\xspace}
\def\Davg{\ensuremath{\Delta_{\Phi \, \textnormal{avg}}}\xspace}
\def\mpadapt{\ensuremath{m_p^{(j)}}\xspace}
\def\mpnecessary{\ensuremath{m_p^{\circ}}\xspace}
\def\qproxy{\ensuremath{q_{t \, (\chi_\parallel)}}\xspace}
\def\PSOL{\ensuremath{P_{\mathrm{SOL}}}\xspace}
\def\qmax#1{\ensuremath{q_{\textnormal{max}}^{\textnormal{(#1)}}}\xspace}
\def\vmax{\ensuremath{\vec{v}_{\textnormal{max}}}\xspace}
\def\Deltagk{\ensuremath{\Delta_{\vec{g}}^{\ast \, (k)}}\xspace}
\def\Deltag{\ensuremath{\Delta_{\vec{g}}^{\ast}}\xspace}

\section{Introduction}

Divertor targets are a key component in magnetic confinement fusion devices - especially in those aiming for burning plasmas \cite{Loarte2007, Krieger2025}.
Power exhaust from the plasma can result in extreme heat loads, and divertor targets need to be designed such that they can tolerate those.
Advanced materials can push the acceptable limit \cite{Zinkle2005}, while driving the plasma in front of the divertor targets into a ``detached'' state \cite{Matthews1995, Krasheninnikov2017, Stangeby2018} can significantly reduce potential heat loads.
Plasma detachment is typically achieved by gas puffing with seeded impurities (N, Ne, Ar, ...) into the divertor volume for power dissipation.

An advantage of stellarator configurations over tokamaks is their potential for continuous operation and improved stability, but this comes at the price of a significantly more complex design.
A particular challenge for stellarators is the toroidal localization of heat loads and the ability to accumulate neutral particles in the divertor volume for efficient removal \cite{Koenig2002}.
High-fidelity numerical models for plasma detachment (such as EMC3-EIRENE \cite{Feng2004}) can guide the design process.
However, they are computationally expensive and are more suitable for design verification from scoping studies based on fast, low-fidelity models.
Even though the latter can not (yet?) capture plasma detachment, they can be leveraged in an optimization loop for minimizing potential heat loads.
While it may be sufficient to reduce the actual heat loads below the limit determined by the material properties (with an additional margin for model uncertainties), minimizing the potential heat loads (i.e. without taking into account power dissipation in the plasma) implies that fewer seeded impurities are required for detachment.

Reconstruction of magnetic field lines from a flux tube mesh provides considerable speedup \cite{Feng2005}.
It is exploited EMC3 for high-fidelity modeling of plasma detchment as well as for low-fidelity heat load approximation.
Recently, an implementation based on an unstructured mesh layout has been integrated into FLARE which is suitable for fast heat load approximation in highly shaped island divertor and non-resonant divertor configurations \cite{Frerichs2025}.
Once the flux tube mesh is constructed, divertor targets can be adjusted as needed within a given magnetic casing (i.e. the first wall, or the boundary of available space for shape optimization).

The present analysis explores the optimization of the divertor target geometry for heat load control.
A brief introduction of the particle swarm optimization (PSO) method is given in section \ref{sec:PSO}.
The PSO method is chosen here because it is easy to implement, does not require information about the gradient of the problem being optimized, and has been empirically shown to perform well on many optimization problems.
Then, in section \ref{sec:Problem}, the specific optimization problem for this analysis is introduced.
Here, rather than jumping into the deep end of stellarator divertor design, the focus is on a toy problem in order to highlight the implications of different parameters.
Specifically, we look at the outer divertor target in the ITER and leverage the axisymmetry of the tokamak configuration in order to reduce the number of free parameters (which is not to say that ITER is a toy or that tokamak divertors are trivial).
The purpose of this simplification is to analyze the potential of PSO for divertor heat load control, which is covered in section \ref{sec:analysis}.


\section{Particle swarm optimization} \label{sec:PSO}

Particle swarm optimization (PSO) is a stochastic optimization method inspired by social behavior of bird flocking or fish schooling \cite{Kennedy1995, Eberhart1995}.
In PSO, a population of candidate solutions (referred to as particles) are iteratively improved by moving around through search space.
The movement is influenced by each particles best known position but also guided towards the global best known position (a variation of the algorithm is instead based on the local neighborhood best).
Each particle $i$ has the following characteristics:

\begin{center}\begin{tabular}{ll}
$\vec{x}_i$:	& the {\it current position} of the particle \\
$\vec{v}_i$:	& the {\it current velocity} of the particle \\
$\vec{p}_i$:	& the {\it personal best known position} of the particle
\end{tabular}\end{center}

The personal best known position $\vec{p}_i$ is either the current position $\vec{x}_i$ or one of its previous values - whichever yields the highest fitness value for that particle.
For minimization of a multi-variate function $f(\vec{x})$, higher fitness is determined by a lower function value.
Furthermore, let $\vec{g}$ be the {\it global best known position} of the entire swarm of particles.
Then, in the next iteration ($t + 1$) of the swarm, the velocity of particle $i$ is updated as follows:

\begin{equation}
\vec{v}_i(t+1) \, = \, w \, \vec{v}_i(t) \, + \, c_1 \, \vec{r}_1 \odot (\vec{p}_i \, - \, \vec{x}_i(t)) \, + \, c_2 \, \vec{r}_2 \odot (\vec{g} \, - \, \vec{x}_i(t)). \label{eq:PSO-Vupdate}
\end{equation}

The parameter $w$ in the first term is referred to as the {\it inertia weight} \cite{Shi1998, Eberhart2000}.
The parameters $c_1$ and $c_2$ are often called {\it cognitive coefficient} and {\it social coefficient}.
One particular combination that will be used later is $w = 0.729$ and $c_1 = c_2 = 1.494$ \cite{Eberhart2000}, but other combinations are possible (within certain limits for stability \cite{Clerc2002, Trelea2003}).
The second and third terms guide the particle towards its personal best and global best known positions with uniform random weights $\vec{r}_1$ and $\vec{r}_2$ which are applied element wise ($\odot$).
The position update is then:

\begin{equation}
\vec{x}_i(t+1) \, = \, \vec{x}_i(t) \, + \, \vec{v}_i(t+1) \label{eq:PSO-Xupdate}
\end{equation}

Particles are initialized with random locations and velocity.
The best known positions $\vec{p}_i$ and $\vec{g}$ are updated when applicable, but not every iteration yields an improvement.
Velocities are clampled to $\pm \vmax$ but with a very generous $\vmax \, = \, \vec{b}_u \, - \, \vec{b}_l$ where $\vec{b}_u$ and $\vec{b}_l$ are the upper and lower bounds of the problem space, respectively.
Particles are allowed to {\it fly} beyond those boundaries, but $f(\vec{x})$ is not evaluated there.


\section{The head load optimization problem} \label{sec:Problem}

Heat loads along the divertor targets and first wall are approximated from the solution of the linearized heat conduction equation $\nabla \, \cdot \, \vec{q} \, = \, 0$ with

\begin{equation}
\vec{q} \, = \, -\kappa_\parallel \, \nabla_\parallel T \, - \, \chi_\perp \, n \, \nabla_\perp T
\end{equation}

where $\kappa_\parallel$, $\chi_\perp$ and $n$ are considered to be fixed model parameters.
More precisely, the thermal conductivy along field lines $\kappa_\parallel \, = \, \kappa_0 \, T^{5/2}$ is based on the physical constant $\kappa_0$ and an input background temperature $T$.
On the target (here: on any surface), the Bohm boundary condition

\begin{equation}
q_t \, = \, \gamma \, T_t \, \Gamma_t
\end{equation}

is applied with $\Gamma_t \, = \, n_t \, c_{st}$ and sound speed $c_{st} \, = \, c_s(T_t)$.
The sheath heat transmission coefficient $\gamma \, = \, \gamma_e \, + \, \gamma_i \, \approx 7$ is applied.
We introduce $\chi_\parallel \, = \, \kappa_\parallel / n$ for convenience in order to relate the impact of cross-field transport to transport along field lines.
The total power into the SOL \PSOL is a scaling parameter to which resulting heat loads can be normalized.
This is the same model as in EMC3-Lite \cite{Feng2022}, but adapted in FLARE for an unstructured flux tube mesh.
Both implementations are based on a Monte Carlo method which relies on fast reconstruction of field lines.
Monte Carlo particles are generated along the inner mesh boundary (typically located just inside the last closed flux surface) and are tracked until they are deposited on the divertor targets or first wall. 

\begin{figure}
\begin{center}
\includegraphics[width=160mm]{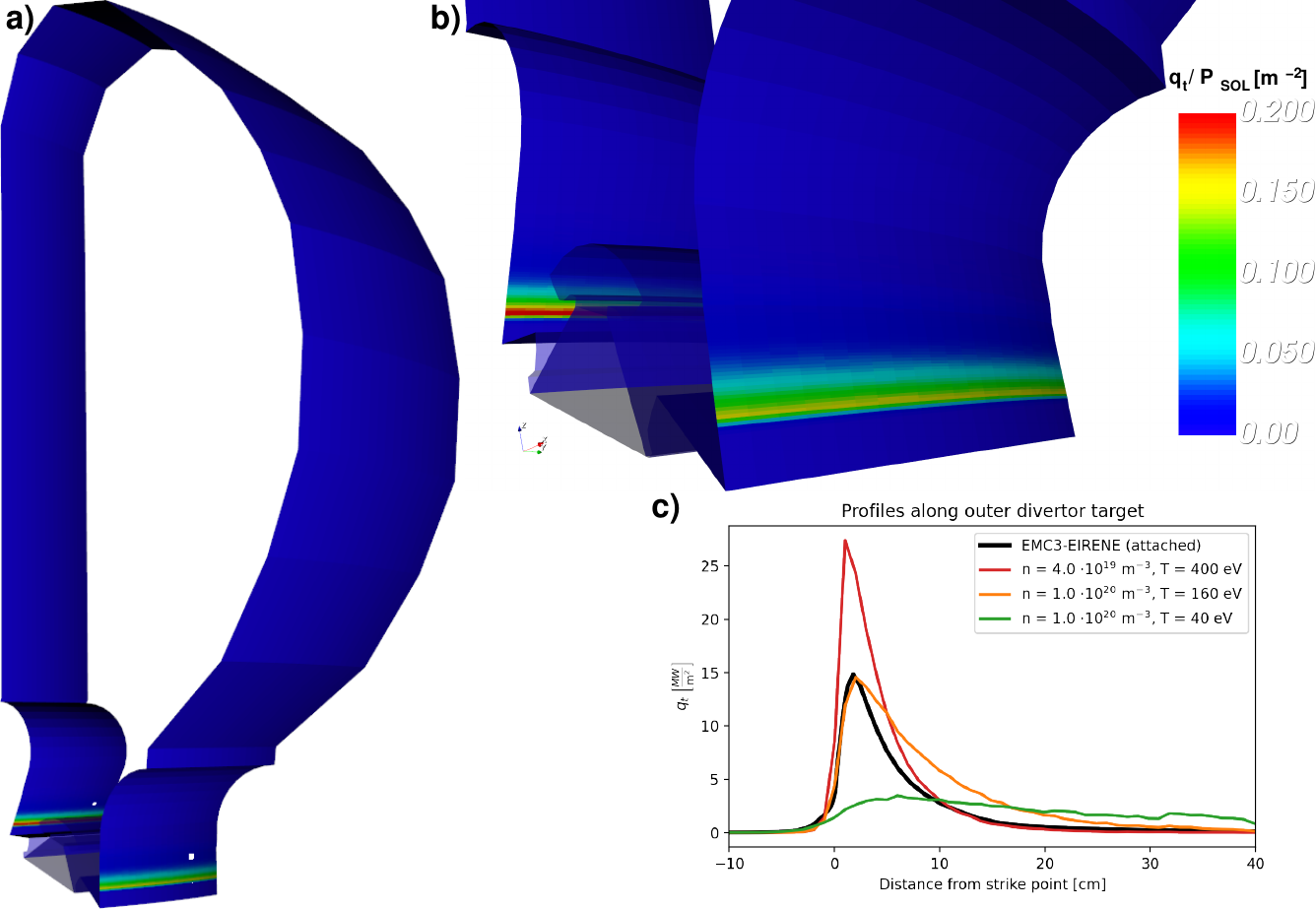}
\caption{(a,b) Simulated heat load distribution along the divertor targets and first wall in ITER. (c) Profiles along the outer divertor target for a few different choices of model parameters. An EMC3-EIRENE simulation (black) is shown for reference.}
\label{fig:Example}
\end{center}
\end{figure}

An example heat load distribution along the divertor targets and first wall is shown in figure \ref{fig:Example} (a,b) where two strike lines can be seen: one on the inner target and one on the outer target.
Model parameters are $n \, = \, 10^{20} \, \meter^{-3}$, $T \, = \, 160 \, \electronvolt$ and $\chi_\perp \, = \, 2.0 \, \meter^2 \, \second^{-1}$.
The simulation has been conducted with $M \, = \, 400000$ Monte Carlo particles and took about 1 minute on a desktop computer with 12 CPUs (using MPI for parallelization).
It should be noted that there are a number of factors determining the run time: the number of Monte Carlo particles $M$, the level of cross-field transport $\chi_\perp$ relative to transport along field lines $\chi_\parallel$, the position of the inner mesh boundary (here: at $\psi_N = 0.99$ normalized poloidal flux), the connection length of field lines in the SOL, and (to some degree) the number of surface patches used to model the divertor targets and first wall.
Figure \ref{fig:Example} (c) shows the heat load profiles on the outer divertor target for a few different choices of model parameters and $\PSOL = 100 \, \mega\watt$ in comparison to an EMC3-EIRENE simulation.
The latter includes Ne seeding at a low level of D${}_2$ gas puffing and is consistent with the corresponding SOLPS-ITER simulation presented in \cite{Pitts2017}.
It can be seen that the heat load proxy based on upstream values (red) is rather peaked while choosing model parameters based on typical downstream values (green) substantially overestimates the broadening of heat loads.
The intermediate case (orange) suggests that a good choice of model parameters should be biased towards upstream conditions.

\begin{figure}
\begin{center}
\includegraphics[width=100mm]{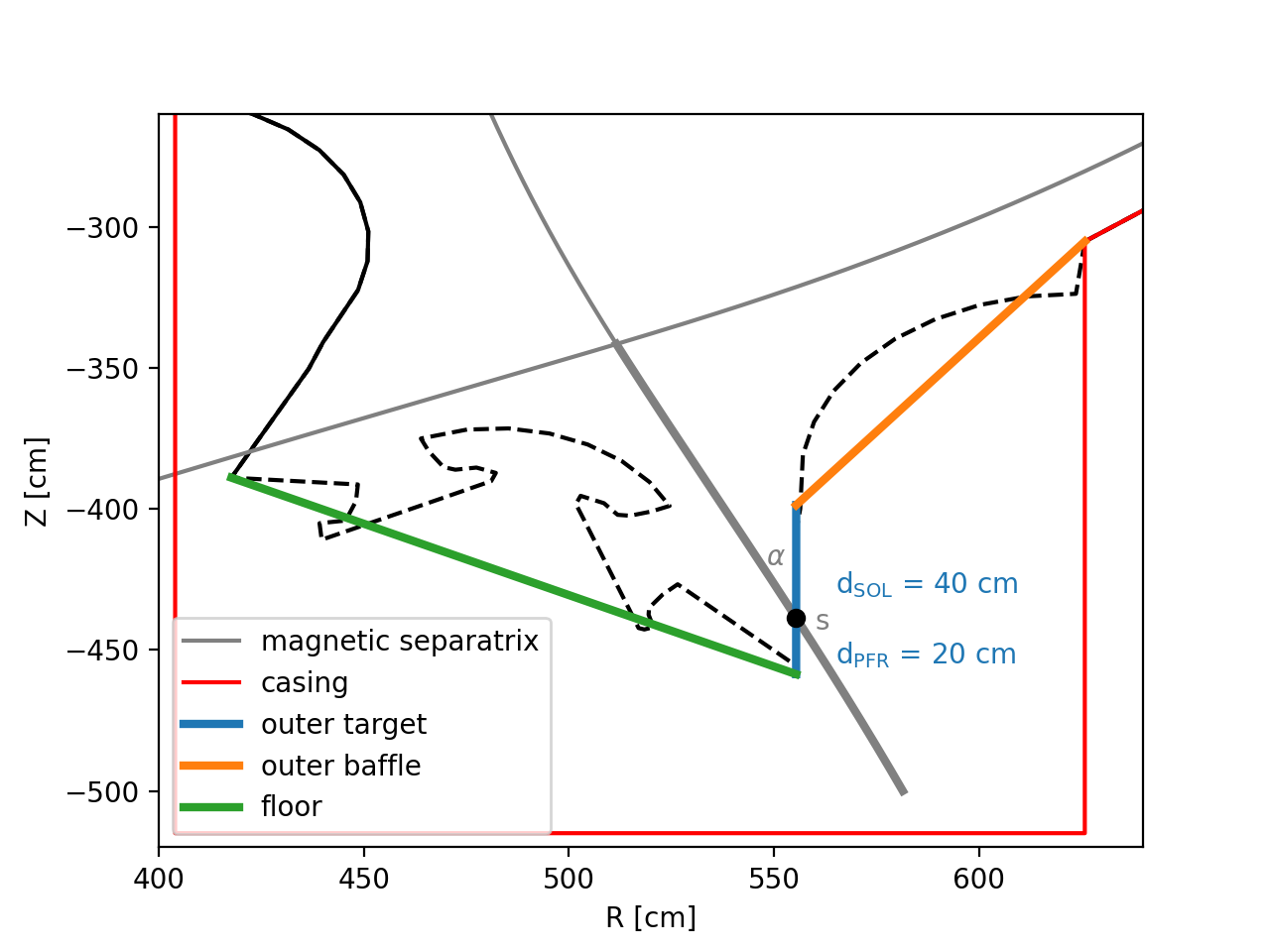}
\caption{Model of the outer divertor target used for heat load optimization. The base point (black dot) is located along the outer divertor leg of the magnetic separatrix (gray). The actual divertor geometry (black dashed line) is shown for reference. The (artificial) boundary for potential configurations is shown in red.}
\label{fig:DivertorModel}
\end{center}
\end{figure}

The ITER divertor includes a dome below the X-point of the magnetic separatrix (as indicated in figure \ref{fig:DivertorModel}.
This is important for efficient pumping of netural particles that are released from the divertor targets after recombination of the incident ion and electron fluxes.
For the purpose of this optimization study, however, we neglect the dome structure and the transport of neutral particles.
Furthermore, we focus on the outer divertor target which we approximate as a plate of fixed length as shown in blue in figure \ref{fig:DivertorModel}.
The base point of the plate (black dot) is anchored along the outer divertor leg of the magnetic separatrix and defined by the variable $s \, \in \, [0, 1]$ going from the lowest point ($s = 0$) up to the X-point ($s = 1$).
The orientation of the plate is determined by the angle $\alpha$ between the surface normal and the magnetic separatrix: i.e. $\alpha = 0 \, \deg$ corresponds to a normal incident of the poloidal projection of the separatrix onto the divertor target (note that this is not the field line incident angle - field line incident is at a grazing angle in toroidal direction).
Positive values $\alpha \in (0, 90) \, \deg$ imply a reflection of neutral particles into the private flux region, as is the case for the actual ITER divertor.
Negative values $\alpha \in (-90, 0) \, \deg$ imply that neutral particles are reflected outwards into the far SOL.
Finally, the divertor volume is closed by a floor connecting the inner and outer targets (indicated in green) and an outer baffle (indicated in orange) connecting to the main chamber first wall.

\begin{figure}
\begin{center}
\includegraphics[width=160mm]{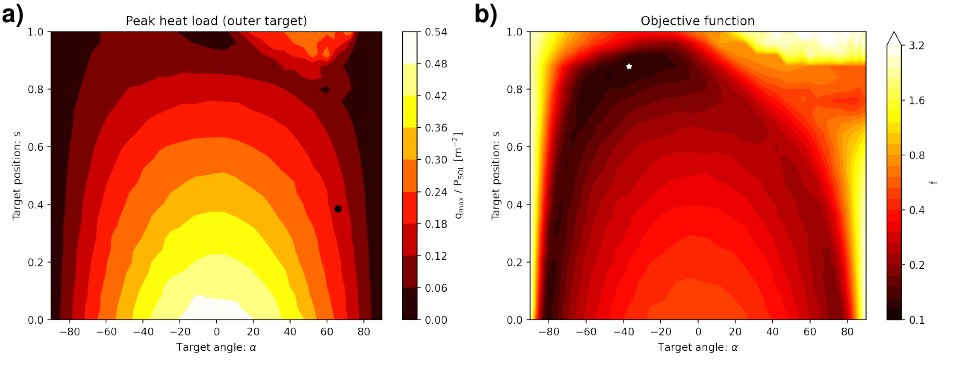}
\caption{(a) Peak heat load \qmax{OT} / \PSOL on outer divertor target for given $(\alpha, s)$. The black dot highlights the actual divertor configuration in ITER. (b) An objective function which penalizes heat loads to the outer baffle and to the floor. The white star marks the position of the global minimum.}
\label{fig:hmax_objective}
\end{center}
\end{figure}

This is a crude approximation of the actual divertor geometry, but the advantage is that it can be described by two parameters such that the entire configuration space can be sampled.
Figure \ref{fig:hmax_objective} (a) shows the dependence of the peak heat load \qmax{OT} on the outer divertor target on $\alpha$ and $s$, and the black dot indicates the actual divertor setup in ITER.
It can be seen that a more grazing incident of the poloidal projection of the magnetic separatrix is more favorable, regardless of the direction in which neutral particles would be reflected (and which are neglected here).
This is to be expected since $\alpha$ is linked to the field line incident angle.
Furthermore, it can be seen that larger $s$ are more favorable for heat loads than smaller $s$.
This is a result of the flux expansion around the X-point.
Obviously, configurations very close to $s = 1$ should be considered with caution (and be excluded) as the target plate may stick into the plasma or shadow the inner target depending on the sign of $\alpha$.

The peak heat load on the outer divertor target should be viewed in context of the other exposed surfaces: reducing \qmax{OT} below the material limit is necessary, but minimizing \qmax{OT} itself can have unintended side effects.
In the following we will consider the objective function

\begin{equation}
f(\alpha, s) \, = \, \qmax{OT} \, + \, f_{\textnormal{penalty}} \, \left(\qmax{baffle} \, + \, \qmax{floor}\right), \qquad f_{\textnormal{penalty}} \, = \, 10 \label{eq:objective}
\end{equation}

which penalizes heat loads to the outer baffle and to the floor and is shown in figure \ref{fig:hmax_objective} (b).
The penalty factor is chosen to reflect different material limits, e.g. $10 \, \mega\watt \, \meter^{-2}$ for divertor targets and $1 \, \mega\watt \, \meter^{-2}$ elsewhere.
It can be seen that configurations with $|\alpha| \approx 90 \, \deg$ are heavily penalized, and so are configurations with $s \approx 1$.
As a result, it is found that there is a main valley of {\it good} configurations along the left side of the plot and a secondary valley on the right side towards smaller $s$.
The latter includes the actual divertor configuratoin in ITER.


\section{Performance analysis} \label{sec:analysis}

\begin{figure}
\begin{center}
\includegraphics[width=160mm]{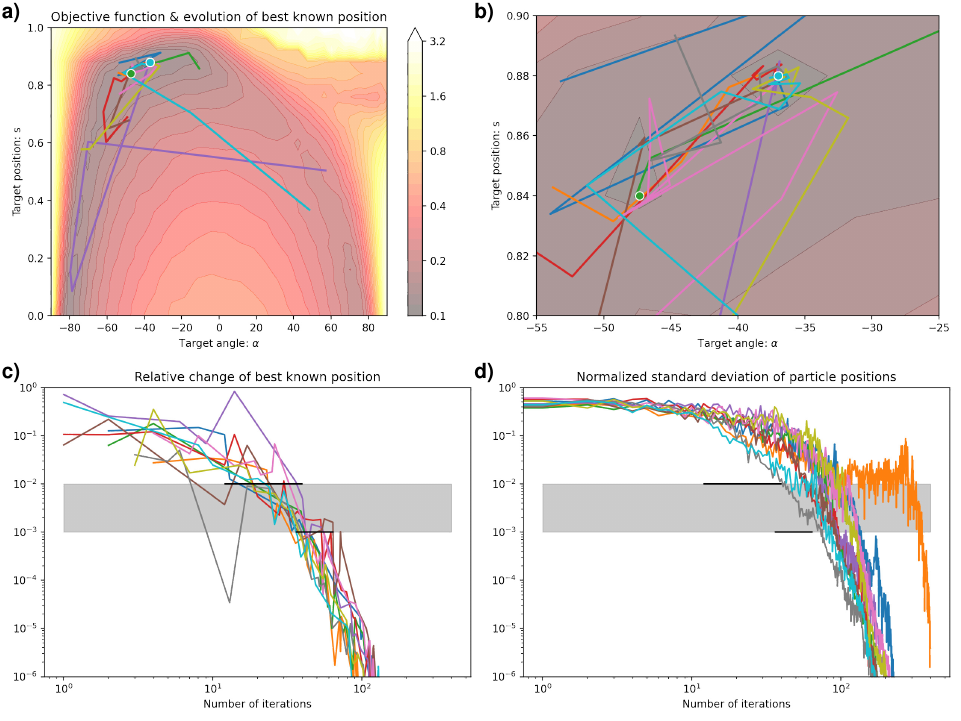}
\caption{(a, b) Evolution of the swarm's best known solution $\vec{g}$ in problem space for 10 swarms of 12 particles each. The final $\vec{g}$ after 400 iterations is highlighted by a colored dot (some points are on top of each other).
(b) Change of each swarm's best known position at the iteration of the update relative to the dimensions of the problem space.
(c) Standard deviation of particle positions within a swarm normalized to the dimensions of the problem space.
}
\label{fig:test1}
\end{center}
\end{figure}

In order to explore the potential of particle swarm optimization (PSO) for divertor heat load control, we take the results from figure \ref{fig:hmax_objective} and construct a continuous objective function via interpolation (using \texttt{RegularGridInterpolator} from Python's \texttt{scipy.interpolate} package).
Snapshots of a particular swarm's evolution are highlighted in \ref{sec:swarm_evolution} where it is shown that a {\it good} solution can be found after only a few iterations.
As PSO is a stochastic algorithm, we will run simulations for a number of swarms for statistical analysis.
Figure \ref{fig:test1} shows the results from 10 swarms with 12 particles each.
It can be seen in figure \ref{fig:test1} (a) that all swarms find a {\it good} solution, but not all swarms find the global minimum.
Instead, one swarm finds a nearby local minimum as highlighted in figure \ref{fig:test1} (b).
Nevertheless, most swarms get within $10 \, \%$ of the global minimum after only a few iterations, and converge to within $1 \, \%$ of their final value (taken after 400 iterations) after 10-40 iterations.

Progress is not always steady: it may take several iterations before a swarm finds a better best known solution.
Let us define $\vec{g}_k$ as the k-th best known position and $i_k$ the iteration number in which it was found.
Then the improvement over the last best known position $\vec{g}_{k-1}$ can be given as the $L^2$ norm

\begin{equation}
\Deltagk \, = \, \|(\vec{g}_k \, - \, \vec{g}_{k-1}) \, \oslash \, \vmax \|
\end{equation}

where the division ($\oslash$) is taken element wise so that the differences are normalized to the dimensions of the problem space.
The improvement \Deltagk at iteration $i_k$ is shown in figure \ref{fig:test1} (c).
The swarms achieve $\Deltag \approx 1 \, \%$ within 12-40 iterations and $\Deltag \approx 0.1 \, \%$ within 36-65 iterations.
Improvements tend to get smaller with advancing iterations, however, a small \Deltag is not a guarantee for convergence.
It is entirely possible that a swarm finds a slightly better solution while it is not yet close to a minimum.
This happens e.g. to the gray swarm in figure \ref{fig:test1} (c) in the 13-th iteration where $\Deltag = 3.4 \cdot 10^{-5}$, but then a better solution is found a few iterations later further away with $\Deltag = 9 \cdot 10^{-3}$.
This is perhaps good enough for divertor optimization given the uncertainty of low-fidelity models, but it is worth noting that stopping the swarm once \Deltag undercuts a given tolerance may lead to premature convergence.

Another quantity that is related to convergence is the swarm's radius. Particles will buzz around their personal best known positions and the swarm's best known position within a volume that tends to shrink around the latter.
Thus, the swarm's normalized radius

\begin{equation}
r \, = \, \| \sigma(\vec{x}) \, \oslash \, \vmax \|
\end{equation}

can be an indicator for the swarm's potential to further explore problem space.
Here, $\sigma(\vec{x})$ is the standard deviation of the particle positions $\vec{x}_i$.
One can even go one step further and define the swarm's energy $E \, = \, \sum_i \, \| \vec{v}_i \|$ in order to avoid the situation where the swarm is momentarily collapsed to a small volume.
However, a weaker - but perhaps sufficient - condition to determine if no further improvement can be expected is the requirement of $r < r_{\textnormal{tol}}$ for two consecutive iterations.
Figure \ref{fig:test1} (d) shows $r$ remains about an order of magnitude larger than \Deltag.
In particular, it can be seen that one swarm remains at a radius of a few $\%$ for a few 100 iterations before it finally coalesces.
At the same time, \Deltag is already below $10^{-6}$.

\begin{figure}
\begin{center}
\includegraphics[width=160mm]{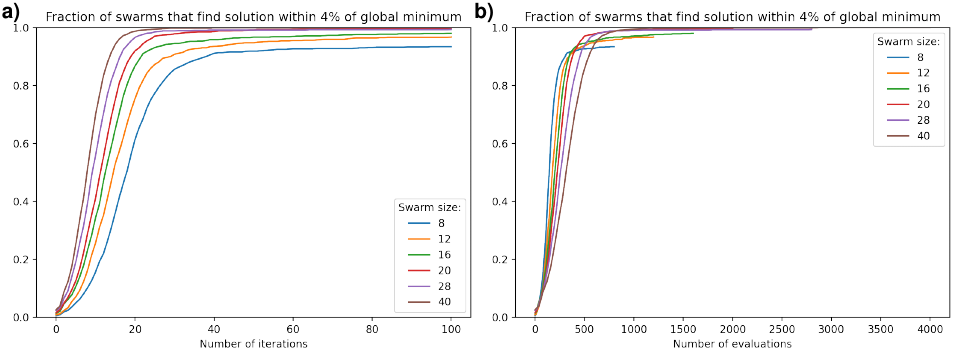}
\caption{Fraction of swarms that reach the global minimum value within $4 \, \%$ accuracy: (a) after given number of iterations, (b) after given number of function evaluations.}
\label{fig:test2}
\end{center}
\end{figure}

Robustness is evaluated in figure \ref{fig:test2} for different swarm sizes.
This time, 1000 swarms are simulated for each swarm size, and the fraction of swarms is evaluated that find the global minimum value within $4 \, \%$ accuracy for a given number of iterations.
It can be seen that a large fraction of swarms reach a good solution within about 20 iterations regardless of the swarm size.
Larger swarms tend to find a good solution within fewer iteration and with higher probability, but this is offset by a larger amount of required evaluations of the objective function.
Figure \ref{fig:test2} (b) shows that larger swarms are not necessarily more efficient.
A swarm size of 12 appears to be working well.

As the swarm is guided by its previously known best positions $\vec{p}_i$ and $\vec{g}$, one aspect to consider is how noisy values of the objective function can impact the optimization results.
Here, the heat load proxy is based on a stochastic method.
Under unlucky circumstances, the peak heat load in a simulation may become the new best known value even if the true value is larger.
This may guide the swarm to converge to a position that is not even a local minimum.
Increasing the number of Monte Carlo particles will mitigate the noise of peak heat load values, but this implies increased computational cost.
Thus, it is important to evaluate the performance of PSO in the presence of noise.

\begin{figure}
\begin{center}
\includegraphics[width=160mm]{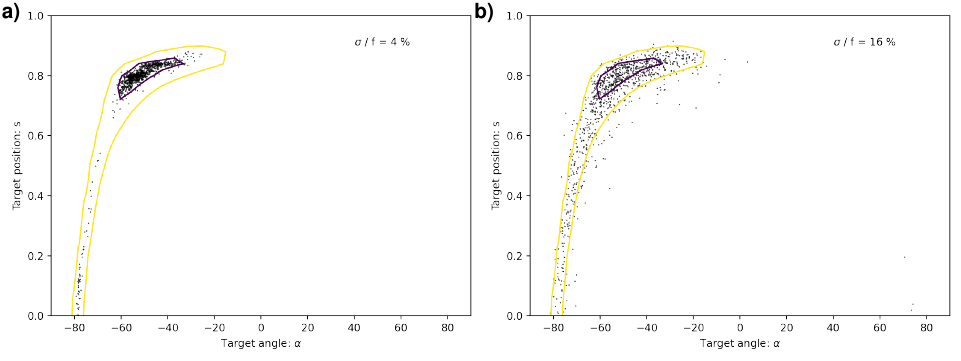}
\caption{Results from 1000 PSO simulations with $4 \, \%$ noise (a) and $16 \, \%$ noise (b). Black dots mark each swarms best known position after 100 iterations. Contour lines for $4 \, \%$ (purple) and $16 \, \%$ (yellow) deviation from $f_\textnormal{min}$ are shown.}
\label{fig:test7}
\end{center}
\end{figure}

The data in figure \ref{fig:hmax_objective} that went into the interpolated objective function already had some noise in it (this is likely the reason for the other, nearby local minimum found in figure \ref{fig:test1} (a,b)).
However, evaluation of the objective function was deterministic.
In the following we will introduce artificial noise to the problem in order to analyze the performance of PSO.
First, though, data values are smoothed with the \texttt{gaussian\_filter} from \texttt{scipy\_ndimage} in order to remove residual noise.
Then, a new objective function $\tilde{f}(\alpha, s)$ is constructed by returning a random sample from a normal distribution with mean $f(\alpha, s)$ and standard deviation $\sigma$.
This noisy objective function is then used in another batch of PSO simulations.
Results from 1000 simulations are shown in figure \ref{fig:test7} for $\sigma / f = 4 \, \%$ and $16 \, \%$.
It can be seen that the swarms no longer converge to the global (or local) minimum, but rather converge to points which lie within the noise range of the minimum value $f_{\textnormal{min}}$.
Target angles of -60 to -35 deg are found to be {\it optimal} at a noise level of $4 \, \%$, and solutions are spread over quite a chunk of problem at a noise level of $16 \, \%$.
It is unclear how this translates to more complex geometries, but it suggests that the potential impact of noise should not be ignored.

\begin{figure}
\begin{center}
\includegraphics[width=160mm]{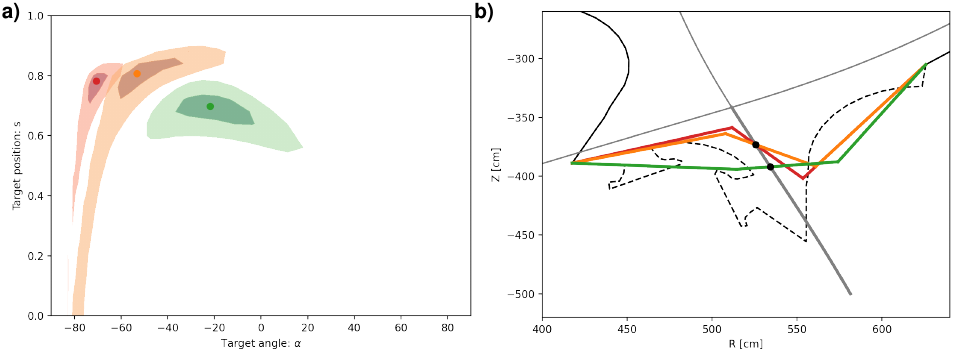}
\caption{Implications of different choices for the background plasma in the heat load proxy simulation with same colors as in figure \ref{fig:Example} (c): (a) configuration space within $4 \, \%$ (dark colors) and $16 \, \%$ (bright colors) of the corresponding global minimum (dots), (b) divertor target and baffle / floor geometry for each optimum.}
\label{fig:Dscan}
\end{center}
\end{figure}

Finally, we circle back to the matter of model parameters for the heat load proxy.
Figure \ref{fig:Example} (c) showed that quite different peak values and profile shapes are found for different approximations of the background plasma.
The implications of those choices are revealed in figure \ref{fig:Dscan}.
It can be seen that a more tangential incident of the magnetic separatrix is preferred for more peaked profiles (red), and that a target location further away is preferred for broader heat load profiles (green).
The resulting shape is quite different from the actual setup, but this is likely caused by the focus on heat loads while neglecting requirements for particle exhaust.
In any case, the candidate solution(s) found by the present procedure would need to be validated by high-fidelity modeling which includes power dissipation by seeded impurities and recycling of neutral particles.


\section{Conclusions}

A procedure for the design of divertor targets for heat load control based on particle swarm optimization has been introduced.
This relies on fast approximation of heat loads with FLARE enabled by reconstruction of field lines from an unstructured magnetic flux tube mesh.
The potential of PSO has been explored based on a simplified divertor geometry for ITER.
Even though that it is not guaranteed that PSO finds the global minimum, the procedure appears to be robust enough to find a solution with good enough accuracy.
However, if this translates to more complex geometries still needs to be verified and compared to other global optimization methods.

The candidate design is found to depend on the choice of model parameters which are necessary for the approximation of heat loads.
It is therefore recommended to generate a few candidate designs for different sets of parameters witin an adequate range, which can then be benchmarked against each other with high-fidelity modeling.
An extension of the procedure to multiobjective optimization which accounts for pumping of neutral particles is in preparation.


\appendix
\section*{Acknowledgments}
This work was supported by the U.S. Department of Energy under award No. P-240001537.

\section{Swarm evolution} \label{sec:swarm_evolution}

\begin{figure}
\begin{center}
\includegraphics[width=160mm]{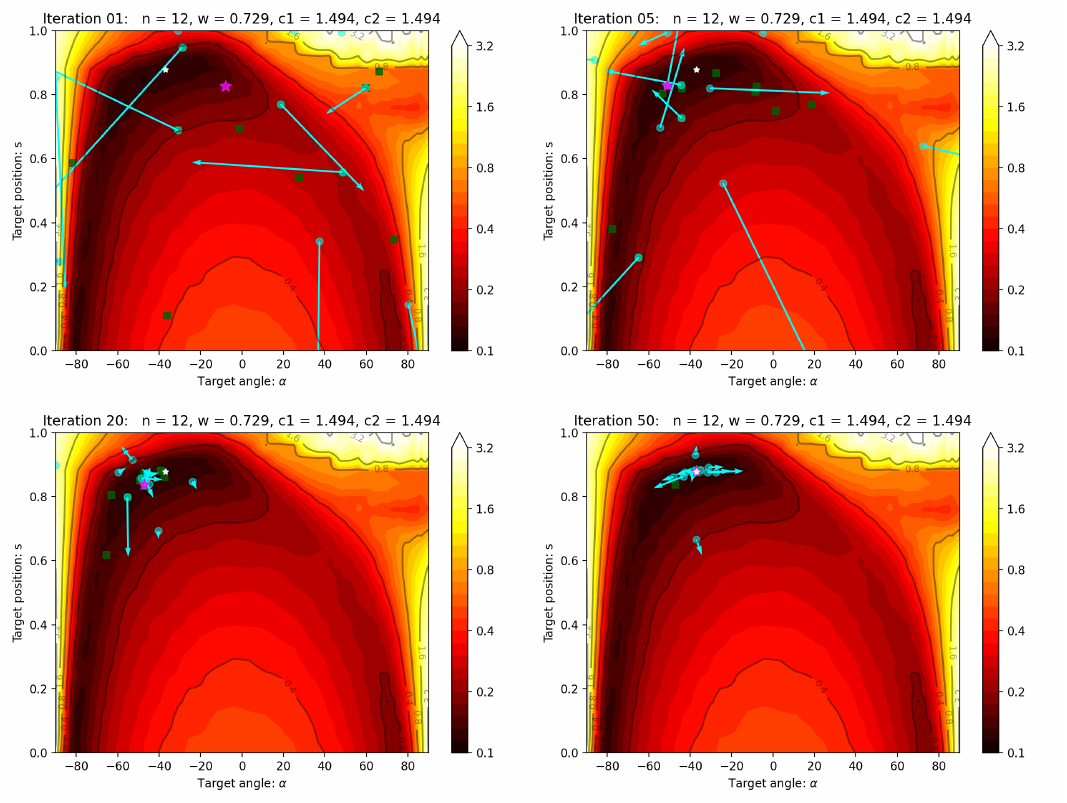}
\caption{Snapshots of the particle swarm after 1, 5, 20 and 50 iterations. Current particle positions and velocities are highlighted in cyan, while personal best known position are marked by green squares. Furthermore, the swarm's best known position is marked by a magenta star and the global minimum is highlighted in white.}
\label{fig:swarm_evolution}
\end{center}
\end{figure}

The evolution of one instance of a swarm of 12 particles is shown in figure \ref{fig:swarm_evolution}.
It only takes a few iterations to find a fair approximation of the global minimum, but this is perhaps related to the low dimension of the problem space.
It can be seen that a good approximation of the global minimum is found after 20 iterations with further refinement after 50 iterations.

\section*{References}
\bibliographystyle{unsrt_doilink}
\bibliography{references}

\end{document}